# Low frequency phase stabilization and phase tuning of an optical lattice with a variable period

P.A. Aksentsev[1,2], V.A. Khlebnikov[1], I.S. Cojocaru[1,3], A.E. Rudnev[1,2], I.A. Pyrkh[1,2], D.A. Kumpilov[1,2], P.V. Trofimova[1], A.M. Ibrahimov[1,2], O.I. Blokhin[1,2], K.O. Frolov[1,2], S.A. Kuzmin[1,2], A.K. Zykova[1], D.A. Pershin[1,3], V.V. Tsyganok[1], A.V. Akimov[1,3]

[1]*Russian Quantum Center, Bolshoy Boulevard 30, building 1, Skolkovo, 143025, Russia*

[2]*Moscow Institute of Physics and Technology, Institutskiy pereulok 9, Dolgoprudny, Moscow Region 141701, Russia*

[3]*PN Lebedev Institute RAS, Leninsky Prospekt 53, Moscow, 119991, Russia*

*email: a.akimov@rqc.ru*

Optical lattices play a significant role in the field of cold atom physics, particularly in quantum simulations. Varying the lattice period is often a useful feature, but it presents the challenge of maintaining lattice phase stability in both stationary and varying-period regimes. Here, we report the realization of a low frequency feedback loop for a tunable optical lattice. Our scheme employs a CCD camera, a computer, and a piezoelectric actuator mounted on a mirror. Using this setup, we significantly improved the long-term stability of an optical lattice over durations exceeding 10 seconds. More importantly, we demonstrated a rapid change in the optical lattice period without any loss of phase. The developed phase stabilization algorithm can be extended to more complicated 2D latices, than just periodic lattice.

## I. INTRODUCTION

Many problems in solid-state physics are difficult to compute and are therefore approached using so-called quantum simulators[1–6]. One example of the systems under investigation is an ensemble of quantum particles in two dimensions. Thin single layer of cold atoms is a cornerstone of the so-called quantum microscopes that give a single particle resolution. These setups allow to investigate such solid-state lattice models as Bose-Hubbard[6–9], Fermi-Hubbard[10–14] and their variations, say with disorder[15]. In addition, two-dimensional dipolar gases have recently attracted significant attention, revealing novel quantum phases such as stripes, droplets, and Berezinskii–Kosterlitz–Thouless superfluidity[16,17]. Besides 2D gases are quite interesting for cold molecular systems[18–20].

To form such a layer, it is convenient to load atoms into a thin disk of light – an optical trap formed by a relatively wide crossing optical beams[21]. Naturally, crossed beams form multiple pancake-like layers of light, and since only one is desired for use, it is advantageous to have an optical lattice in which the distance between these layers can be dynamically changed. Due to the structure of such an optical lattice and its variable interlayer spacing, devices that generate this kind of controllable lattice have been called optical accordions[22–28]. The flexible structure of this trap allows nearly all atoms from a 3D ultracold atomic cloud – typically 15–30 μm in size – to be loaded into a single lattice layer. The cloud can then be compressed into a quasi-2D system, with one dimension reduced to the order of a micron, by changing the lattice period.

Figure 1 represents the concept of a variable distance between optical layers, which corresponds to the maxima of an interference pattern formed by two beams intersecting at an angle. If the angle between the two beams forming the optical lattice is $2\theta$ then the intensity in the interference region is:

$$\begin{aligned}
&I(x) \propto 1+\cos(\Psi) \\
&\Psi = 2kx\sin\theta+\psi = \frac{2\pi}{\Lambda}x+\psi \\
&\Lambda = \frac{\lambda}{2\sin\theta} \\
&k = \frac{2\pi}{\lambda}
\end{aligned} \tag{1}$$

where $k$ is the wave number of light with the wavelength $\lambda$, $\psi$ is the relative phase shift of the interfering beams resulting from their optical path difference, and $\Lambda$ is the period of the optical lattice. The optical lattice period is changed by varying the intersection angle $2\theta$ of the crossed beams, which is in turn controlled by the distance $D$ between the beams before the lens (see Figure 1):

$$\Lambda = \frac{\lambda}{2}\sqrt{1+\left(\frac{2f}{D}\right)^2}\,, \tag{2}$$

where $f$ – is the lens focal length; strictly speaking, the last equality holds exactly only for a thin lens. In our case, $\lambda = 1064$ nm light is used, and a 2-inch $f = 49$ mm aspheric lens focuses the beams which have $1/e^2$ waists on the order of 1 mm near the lens.

The optical lattice overall phase $\Psi$ instability, or equivalently node position instability, can occur for several reasons (1). The relative phase shift $\psi$ of the interfering beams and the intersection angle $\theta$ can change due to vibrations of any reflective surface along the beam's path. Additionally, the light wave number $k$ may fluctuate. These variations shift the phase as:

$$\Delta\Psi = \Delta\psi + 2x sin(\theta)\Delta k + 2kx cos(\theta)\Delta\theta . \quad (3)$$

In turn, lattice phase instability can lead to atomic losses due to several effects:

0. Parametric resonance heating[29,30], which occurs when the lattice trap frequency is $2n$ ( $n$ – integer) multiple of the phase instability frequency, causing an exponential increase in the oscillation amplitude of atoms.

1. Loading losses occur if there is a mismatch between the lattice maximum and the atomic cloud center while loading atoms into the trap. This scenario is indicated in Figure 1 as Problem 1;

2. Losses due to large phase instability of an optical lattice during lattice tuning. If the lattice moves faster than atoms can follow, atoms will be lost. This scenario is indicated as Problem 2 in Figure 1;

3. Losses due to drift of phase of an optical lattice. If the lattice drifts at a sufficiently fast rate, atoms may remain on the side of the trapping potential, as shown in Problem 3, in Figure 1. At this point, the effective trapping depth is reduced, leading to faster atom leakage. Additionally, continuous drift can cause atoms to leave the focal depth of the microscope used for observation. The optical axis of the detection system is aligned along the grating direction (see Figure 1). For our purposes, the allowable mismatch $\Delta_f$ between the position of the optical lattice layer containing atoms and the focus of the atomic detection module is $\pm 0.9$ μm. For a larger mismatch, the image sharpness will be insufficient to resolve individual sites of the 2D lattice (see above). Therefore, for the largest achievable lattice period of 12 μm, the phase must be stabilized with an accuracy of at least $\frac{2\pi\Delta_f}{\Lambda} = 0.47$ rad.

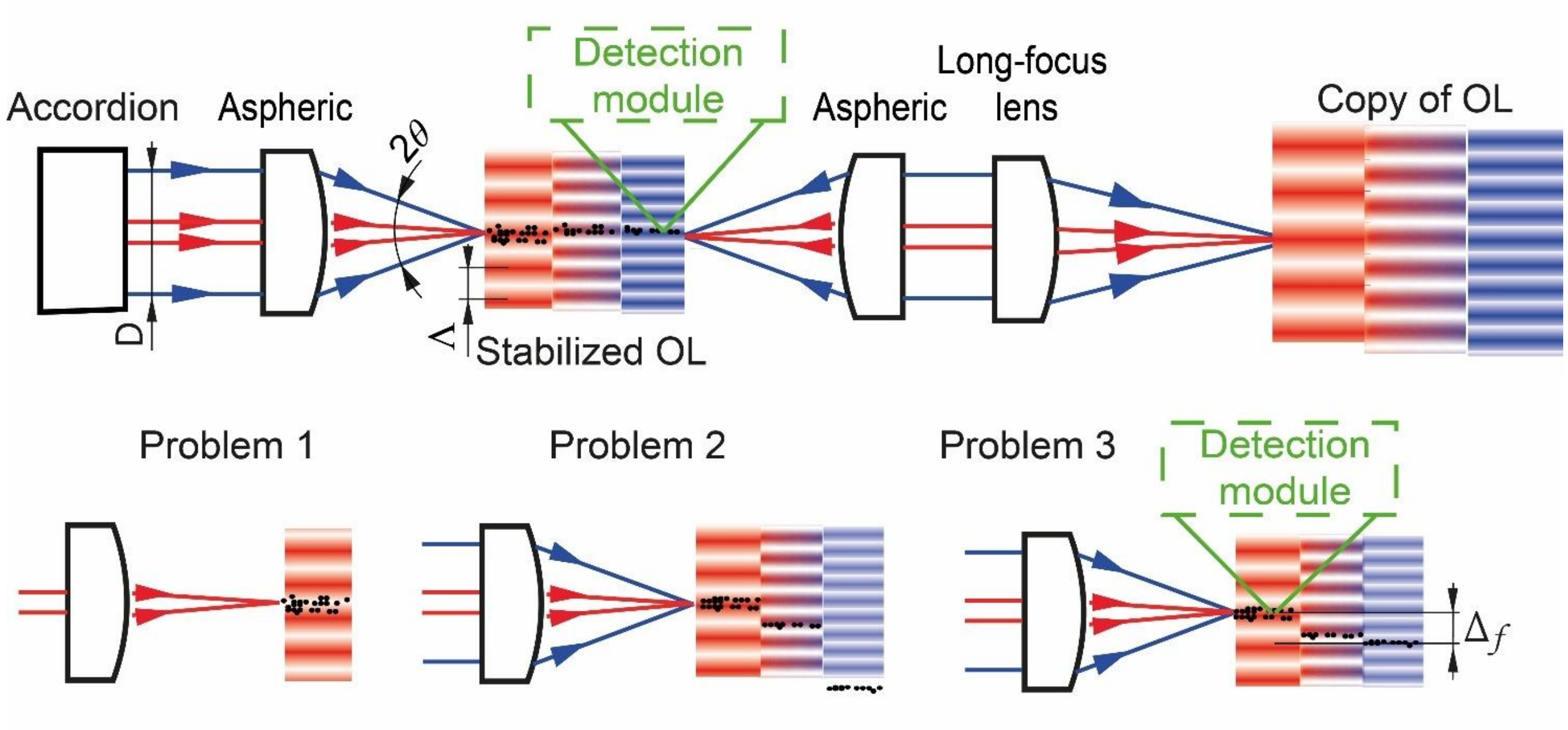

Figure 1. The scheme of the optical lattice (OL) trap and its magnified copy, which is used to stabilize the lattice phase, are shown in the first row. The colors of the beams indicate the relationship between the distance between the incoming beams and the period of the corresponding lattice, which results from their interference. Possible issues that may arise while manipulating atoms in the optical lattice trap are illustrated in the second row.

Since there is no way to read the phase of the optical lattice in real time without destroying the atomic ensemble, it is convenient to use a duplicate image of the trap, as shown in Figure 1. In such a system, stabilization can be performed using this image, which may also be magnified for convenience. The phase of the optical lattice $\Psi$ to be stabilized can be detected using various methods, for example with an aperture with photodiode[31], an array of photodiodes[32], or a CCD camera. As demonstrated in[33], the phase change in the duplicated lattice is identical, up to a sign, to the phase change in the atomic lattice. Indeed, the lens configuration used to create the duplicated lattice forms an enlarged and inverted image of the optical accordion lattice. For the experiments below, a CCD camera was chosen for phase detection, as it provides the most straightforward method to identify the optical lattice phase in the case of a lattice with a variable period.

## II. EXPERIMENTAL SETUP

The contrast of the optical lattice depends on the polarization of the interfering beams. The highest contrast is achieved when both beams have linear polarizations perpendicular to the plane of Figure 2. For this reason, schemes using one[24] or two[22] beamsplitters are used.

Two alternative optical schemes, shown in Figure 2, were investigated experimentally. These schemes differ in the translation element used. The setup is similar to the one presented in[6]. The laser light enters the scheme by reflecting off a mirror and passing through a $\lambda/2$ waveplate (WPH10M-1064). A portion of the beam, with polarization perpendicular to the plane of Figure 2, is reflected by the first polarizing beamsplitter (PBS25-1064-HP), while the remaining portion is transmitted through both beamsplitters. The transmitted beam, after reflection from the mirror and double-passing a $\lambda/4$ waveplate (WPQ10M-1064), acquires linear polarization perpendicular to the plane of Figure 2, causing it to be reflected at the second polarizing beamsplitter. A custom aspheric lens (focal length of 49 mm) is used to form the optical lattice targeted for stabilization. The rotation of the $\lambda/2$ waveplate allows control over the power ratio between the two portions of the laser beam.

To replicate the optical lattice, an identical aspheric lens paired with a 1000 mm telephoto lens (LA4663-AB-ML) was used. This magnification system provides sufficient resolution (>2 pixels per lattice period), enabling the CCD camera Mars5000s-35μm (pixel size: 3.45×3.45 μm) to accurately resolve both the lattice period and phase across all tested period values. The 1 inch lens here limits the lattice period in the range of 2 to 12 μm. The feedback loop has following elements after the camera: the camera is conneted to the computer via USB port (5 Gbps speed for data transmission), CPU processor processing photo with efficient rate of about 1 kHz and calculate error signal, next it generated by the National Instruments PCIe-6363 board is 10 kHz and then fed to the piezo actuator driver MDT 693B with bandwidth of 3 kHz.

The optical lattice period is adjusted in the first scheme by shifting the first mirror in the setup (Figure 2a), and in the second scheme by shifting the paired beamsplitters (Figure 2b). In both cases, the moving element is translated using the DDS050/m translation stage by Thorlabs.

To visualize the phase instability, the camera was set to continually capture images of the duplicate of the optical lattice during the operation of the moving part, which altered the distance between the optical beams. By combining the central columns of pixels from each image, the phase dynamics could be tracked over time. The geometry demonstrated in Figure 2b performs poorly: the lattice compresses, accompanied by significant phase fluctuations.

In contrast, if the mirror reflecting the light onto the beamsplitters is moved along the incident beam axis (Figure 2a), transverse vibrations of the optical elements do not contribute to optical lattice phase instability. This results in a more stable accordion operation. This configuration did not exhibit Problem 2 and showed only a weak manifestation of Problem 3, allowing lattice compression even without active stabilization. However, Problem 1 remains to be addressed in this case.

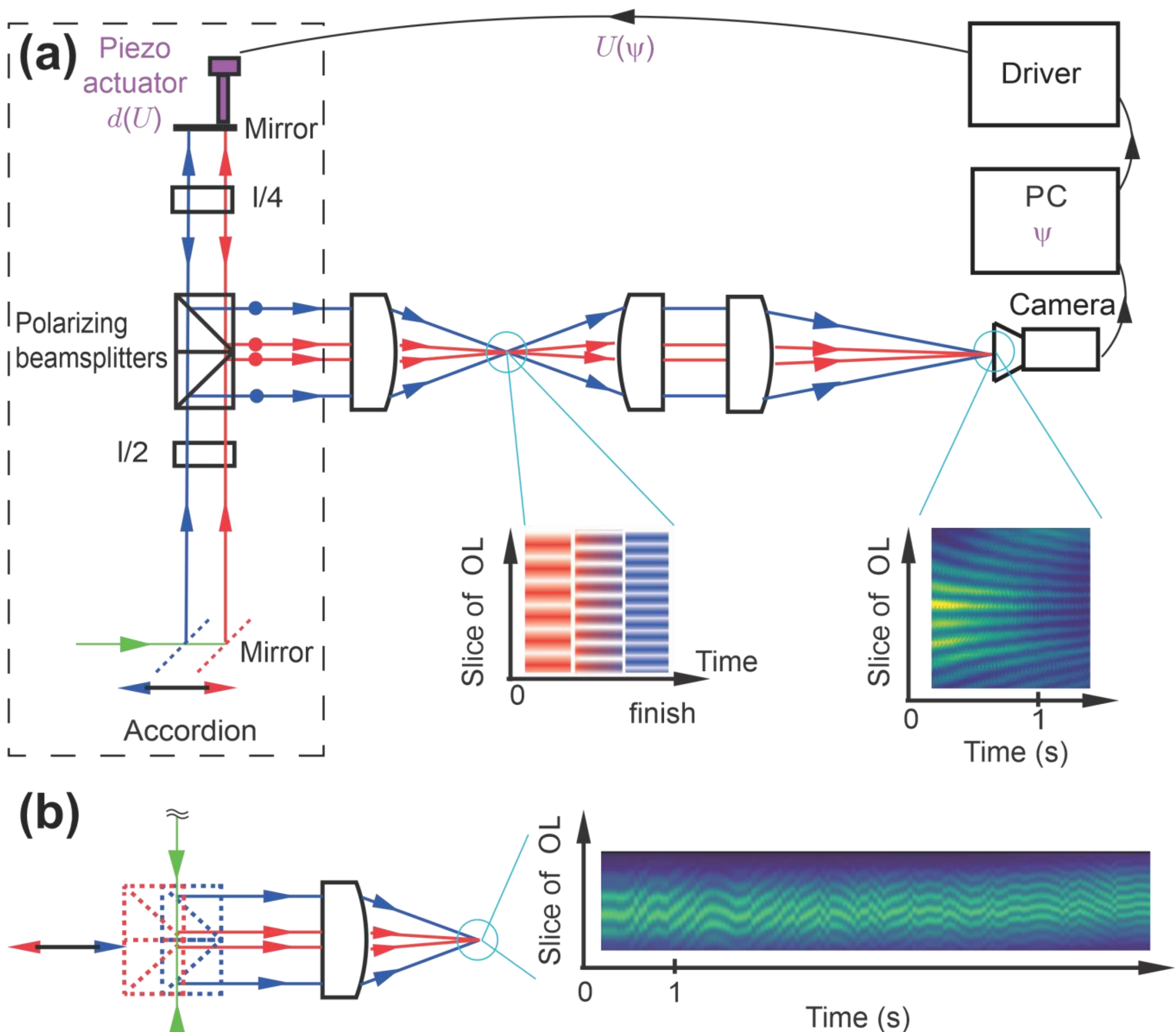

Figure 2. a) Optical accordion implementation using a moving mirror, which provides stable phase behavior during operation. On the right, a composite image (splice of slices) from a series of optical lattice photos illustrates the accordion compression during mirror movement. b) When beam splitters are translated, the phase becomes highly unstable due

to vibrations transverse to the final optical axis. On the right, a composite image during beam splitter movement.

## III. STABILIZATION METHOD

To stabilize or simply analyze the current state of the optical lattice, the following procedure is implemented. The camera periodically captures images (see inset in Figure 3a) of the copy of the optical lattice and sends them to a computer. The computer sums the pixel values along the interference lines, then pads the resulting data array with zeros until the size $s$ of the array becomes a number equal to a power of 2. A fast Fourier transform (FFT) is then performed. The value of frequency $p$ corresponding to the maximum in the magnitude of the Fourier spectrum (green line in Figure 3a) provides the period of the analyzed optical lattice:

$$\Lambda = (s/p)(px/mag), \tag{4}$$

with $px$ is the camera's pixel size and $mag$ is the magnification of the aspheric–telephoto lens pair that forms the copy of the optical lattice (see Figure 2a). The imaginary part $\psi_F$ (blue line in Figure 3a) of the Fourier series for the frequency $p$ gives, up to a constant shift, a quantity associated with the phase $\Psi$ of the optical lattice:

$$\Psi = -\psi_F + const. \tag{5}$$

We take $const = 0$. To approve that statement the system were simulated in Zemax optic studio non sequential mode[34] The changes in phase can differ $\pm 2\pi$ because $\psi_F$ is constrained to values within an interval of length $2\pi$ (one period), while the optical lattice layer containing atoms may shift by distances larger than one lattice period. If the differences $\Delta\psi_F$ between adjacent measurements exceeds $\pm 3$ rad, we subtract or add $2\pi$ to the last obtained phase value so that discontinuities in $\psi_F$ are eliminated.

Based on the extracted values $\psi_F$, the voltage applied to a piezo actuator of the movable mirror (see the left top corner of Figure 2a) is adjusted by $\Delta U_c$ as:

$$\begin{aligned} e_i &= b - \psi_{F_i} \\ \Delta U_c &= P \cdot e_c + I \sum_i^c e_i \end{aligned}, \tag{6}$$

thus changing the optical path difference between the interfering beams or equivalently their relative phase $\psi$ and correspondently the overall phase $\Psi$ (see (1) and (3)). In (6), $e_i$ is the phase error at the $i$-th step where all $i$-th steps precede the current step $c$; $b$ is the preset lattice phase offset; $P$ and $I$ are the proportional and integral coefficients, respectively, which were tuned using the Ziegler–Nichols method[35]. Thus, while compensating for phase drift, the total applied voltage evolves as:

$$U_c = U_0 + \sum_i^c U_i \, , \tag{7}$$

With the initial value $U_0$ chosen near the center of the piezoelectric actuator's dynamic range.

## IV. DRIFT AND NOISE ESTIMATION

Eliminating drift requires a slow stabilization channel for the optical lattice. One possible cause of such drift is a change in ambient temperature. The correlation between lattice phase and temperature was measured experimentally for a non-stabilized, stationary optical lattice (Figure 3b). The measurements were conducted using a laser that had already been running for 5 hours to ensure the system had reached its steady state. Data on the slow phase drift were collected every minute, showing a phase shift of approximately 1 rad per hour. Simultaneously, the room temperature was recorded using a DS18B20 thermometer with a 1 minute sampling interval (Figure 3b). The full measurement lasted 11 hours. The correlation coefficient between temperature fluctuations and lattice phase drift over the entire measurement period was found to be 0.5 −, indicating moderate correlation between the two processes (see the inset on Figure 3b). However, this also shows that eliminating temperature instability alone is insufficient to fully compensate for lattice drift – active stabilization is therefore required.

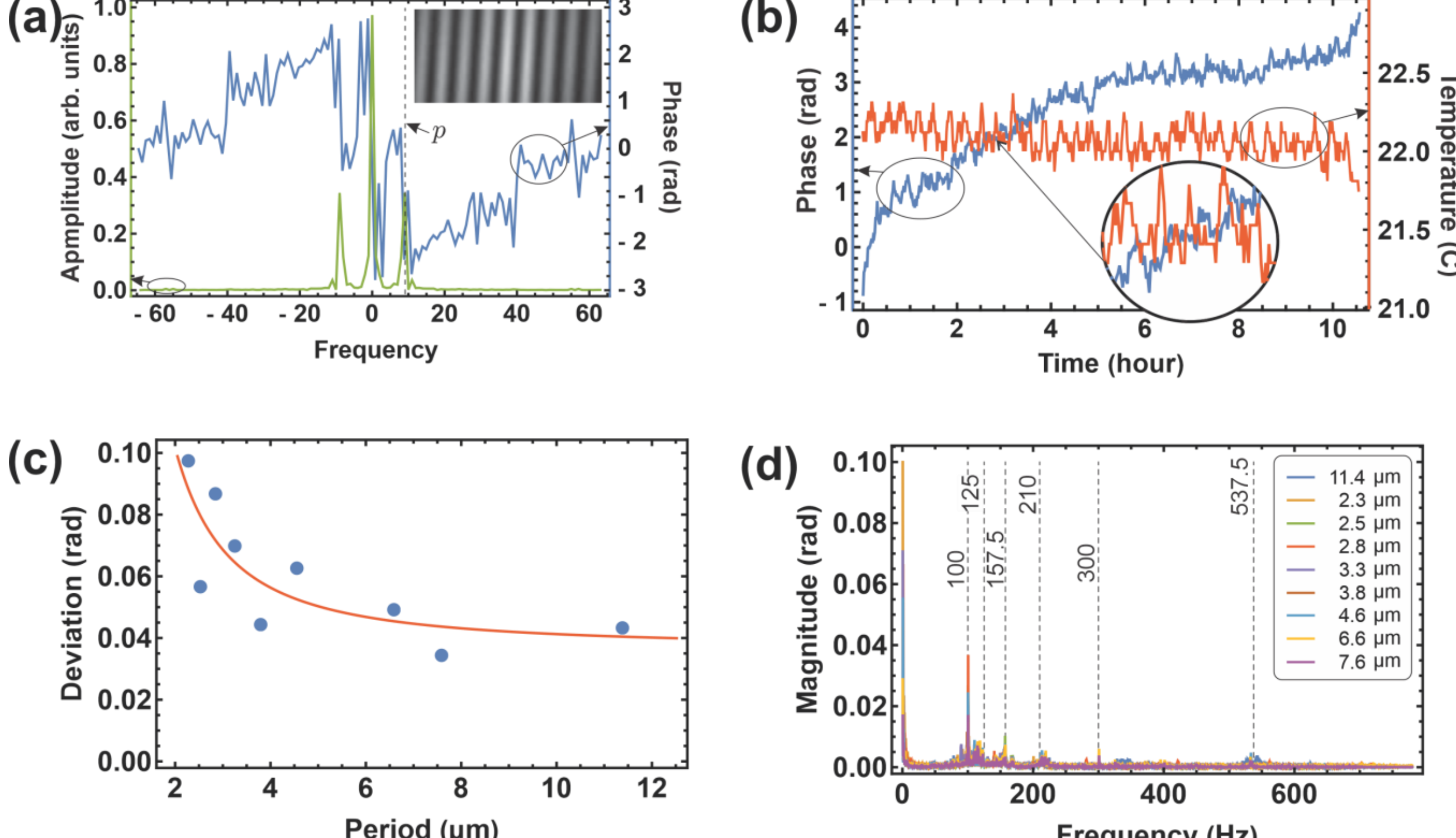

Figure 3. a) Spatial spectrum of the lattice image and its phase. The inset demonstrates the image used for spectrum calculation. b) Slow evolution of phase and temperature. Phase drift was measured every minute. The inset highlights partial anti-correlation between abrupt changes of the ambient temperature and optical lattice phase. c) Standard deviation of short-term phase noise for different periods. The red line represents the fit with equation (8). d) Optical lattice phase noise spectrum for different periods.

To determine the magnitude of high-frequency phase fluctuations, we conducted measurements similar to those used for analyzing slow phase changes, but at the maximum allowable frame rate of the camera – 0.65 milliseconds per frame. The data obtained were post-processed using the method described in the previous section. High-frequency measurements were performed for various lattice periods over a total duration of 3.5 seconds. From this data, the standard deviations of the phase noise were calculated. For all lattice periods investigated, the standard deviations were significantly lower than 0.47 rad – the threshold set by the detection module. According to (3), phase fluctuations are expected to decrease with increasing lattice period. Let us now rewrite (3) to express the phase change as a function of the optical lattice period only:

$$\Delta\Psi = a_1 + \frac{a_2}{\Lambda} + a_3\sqrt{1-\left(\frac{\lambda}{2\Lambda}\right)^2}\,. \tag{8}$$

where $a_1, a_2, a_3$ are constants. The obtained standard deviations of the phase fluctuations were fitted to this functional dependence, as shown in Figure 3c.

To determine the frequency spectrum of the phase noise, we performed a Fourier transform of the rapid phase fluctuation measurements (Figure 3d). The resulting phase noise spectrum reveals several resonance frequencies, which were found to be independent of the lattice period. Since all resonance frequencies are identifiable, they can be avoided – for example, by tuning the oscillation frequency of atoms in the trap via light intensity. As a result, the only remaining instability that must be actively compensated is the slow phase drift.

Although the phase noise spectrum contains components up to 540 Hz, the feedback loop is limited by its response crossover frequency, which is around 100 Hz, according to the measured phase frequency response[36] (Figure 4a). After the phase delay in the phase frequency response exceeds the value of pi, the feedback begins to work as a generator and increases the error instead of minimizing it.

We also compute the two-sample or Allan variance[37,38] $\sigma_y^2(\tau)$, which is widely used to characterize the time stability of a signal generator operating at a frequency $y(t)$. The Allan variance is defined as the variance of the average frequencies measured over two adjacent time intervals of duration $\tau$. The minimum of the Allan variance indicates the time interval $\tau$ over which the frequency deviation between the start and end of an interval is minimized, on average, regardless of the starting point $t$. Moreover, the behavior of the Allan variance as a function of $\tau$ is often associated with specific types of noise sources in a setup. We define $y$ as:

$$y(t) = \left(\Psi(t)_\tau - \Psi(t+\tau)_\tau\right)/\tau \ , \tag{9}$$

so the Allan variance is:

$$\sigma_y^2(\tau) = \frac{1}{2}\overline{\left(y(t+\tau)_t - y(t)_t\right)^2} \tag{10}$$

and is presented in Figure 4b for both stabilized (orange curve) and non-stabilized (blue curve) static optical lattice. The decreasing part of $\sigma_y^2(\tau)$ follows a $\tau^{-2}$ dependence. For a generator, such behavior indicates the presence of white phase noise in the signal[38]. The transition from decreasing to increasing $(\sim\tau)$ in the Allan variance of a non-stabilized optical accordion manifests the onset of random fluctuations in $y(t)$ on time scales of an hour or more. This behavior is often caused by ambient temperature variations, vibrations of optical elements, or noises on sound frequencies[38,39]. For time scales of tenths of seconds and longer, the stabilization reduces the Allan variance by an order of magnitude and effectively eliminates phase drift (compare Figure 3a and Figure 4c).

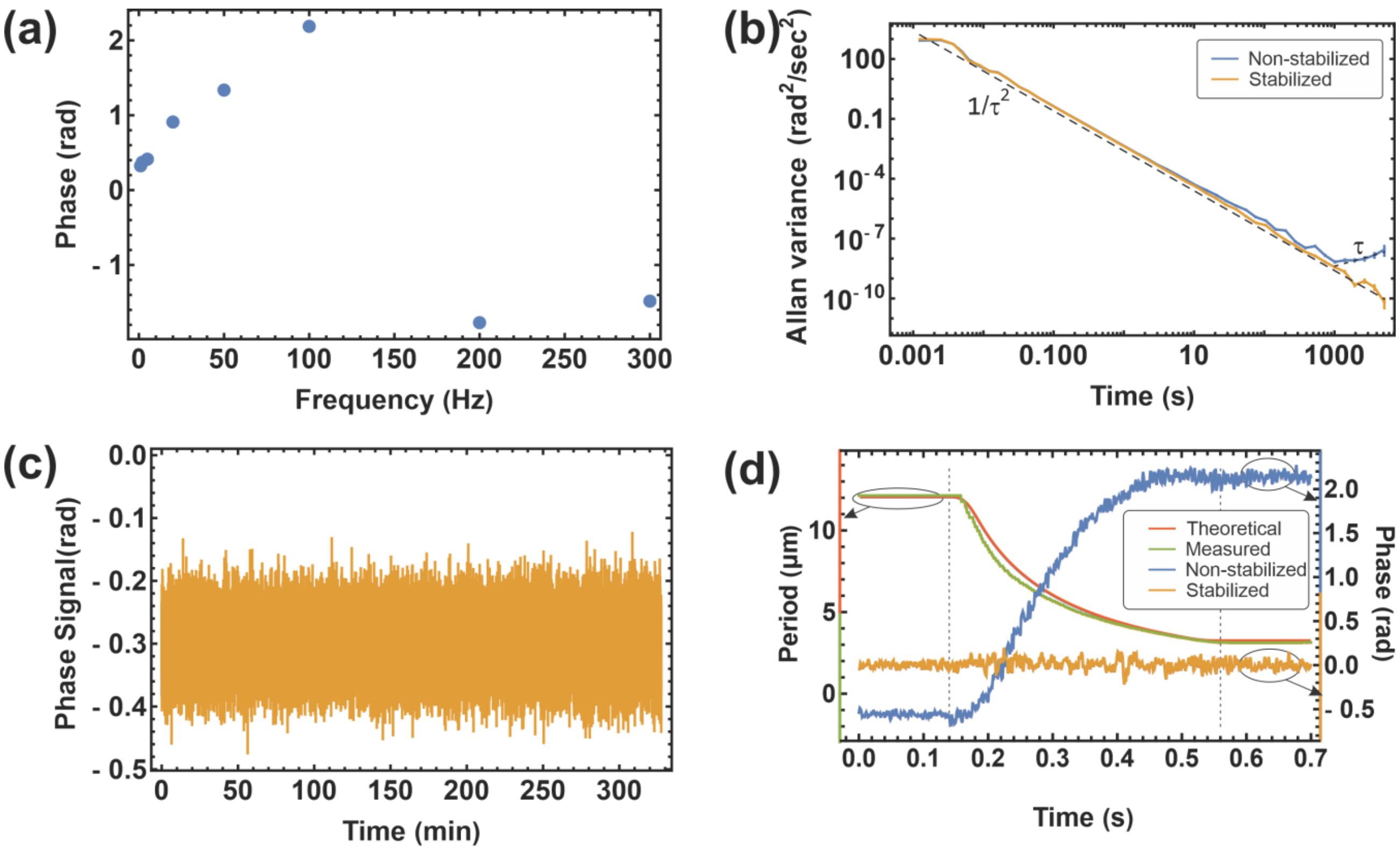

Figure 4. a) Phase frequency response of the feedback loop. b) Allan deviation of the optical lattice phase. Blue curve – non-stabilized phase, orange curve – stabilized phase. c) Optical lattice phase with the feedback loop operating at 800 Hz. d) Phase stabilization during optical accordion operation. The green curve shows the optical lattice period measured via Fourier analysis of the images. The red curve represents the theoretical lattice period calculated using Equation (2). The blue curve shows the phase of the optical lattice under an unstable motorized device regime (see text). The orange curve shows the stabilized phase under the same conditions. Dashed lines indicate the start and stop of the OL period change.

## V. OPTICAL LATTICE PHASE STABILIZATION

If one closes the stabilization feedback loop, the drift is considerably suppressed. The proportional and integral coefficients of the feedback loop were set using the Ziegler-Nichols method. We repeated the long-duration measurement of the optical lattice phase, this time with locked feedback. The loop operated at 800 Hz. The feedback system performed continuous phase reading and applied the corresponding control voltage to the piezo actuator. The results, recorded over 5 hours, are shown in

Figure 4c. The feedback effectively compensates for the slow drift, maintaining the phase within the acceptable range of ±0.47 rad.

The stabilization system was also tested during the operation of the optical accordion to demonstrate its effectiveness in compensating for Problem 3. Figure 4d shows both the stabilized and non-stabilized phase behavior of the optical accordion during operation. Without stabilization, the phase change at the high speed of the motorized stage reached 2.5 radians (blue curve), whereas with locked feedback, the phase deviation was reduced to 0.04 rad (orange curve). The lattice period at each step was calculated using FFT analysis of each captured image (green curve). The theoretical curve of the lattice period (red curve), evaluated based on equation (2) based on the known displacement of the motor in time, shows good agreement with the measured lattice's period data. These results confirm that stabilization enables reliable operation of the optical accordion even in fast regimes. It is worth noting that at moderate motor velocities, the phase remains sufficiently stable (Figure 2a) and the accordion can be operated without stabilization.

The effect of the feedback loop could be also understood from noise spectrum, presented at the Figure 5. For stable configuration the feedback only compensates slow drifts happening on hundreds of milliseconds scale and slower since the loop is already quite stable (Figure 5a). But in the case of motion of the stage, the optical lattice is less stable and therefore feedback has to capture faster changes, and is capable to compensate it up to around 200 Hz.

The speed with which lattice could be compressed is limited by both amount of phase change for unestablished lattice compression (see Figure 4) and effective bandwidth of the feedback of around 200 Hz (see Figure 5b). The 200 Hz bandwidth is below expected, but matches measured above phase frequency response (Figure 4a). The low feedback cutoff can be related to intrinsic delay in the camera output. Camera presumably does shoot at the specified rate, but returns frames with some delay, which reduces bandwidth of the feedback loop.

In case of the described experiment the stage was moving on total of 6 mm. Its maximum speed is $20\ \mathrm{mm/s}$, and maximum acceleration is $600\ \mathrm{mm/s^2}$. Thus, whole compression or expansion of the lattice can happen no faster than $6/20 \approx 0.3\ \mathrm{s}$ with phase instability having standard deviation of 0.064 rad and peak to peak variation of about 0.4 rad as it is shown at Figure 4d. The 0.45 radians variation corresponds to 5 % level of the optical lattice intensity maximum width, which is similar to the resolution limited criteria, which was taken as an acceptable here. If one can tolerate large phase variations, no doubt the compression can be done somewhat faster with the larger phase instability of the lattice.

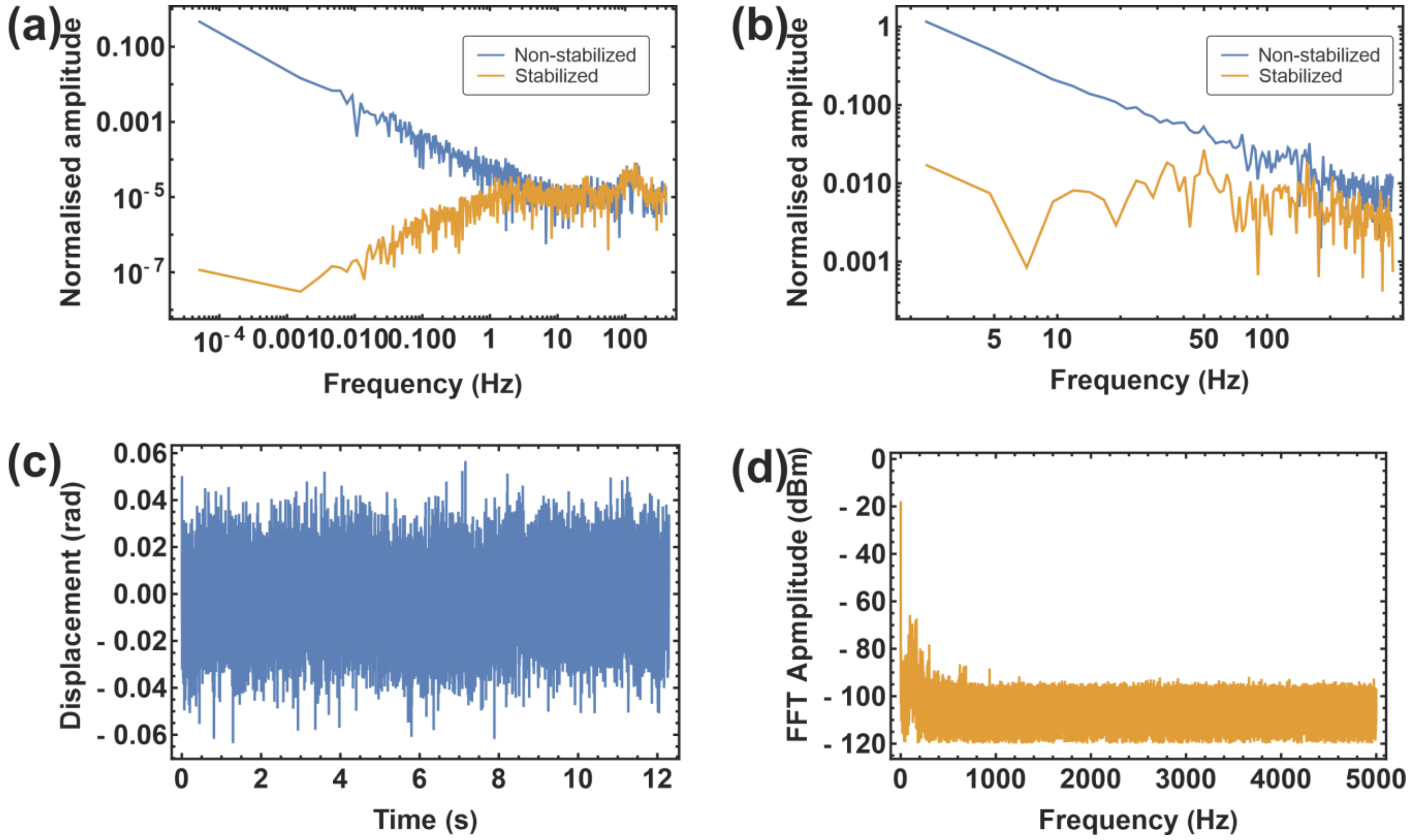

Figure 5. a) Noise spectrum of the feedback signal for locked and unlocked feedback loop in static configuration, b) Same in case of stabilization during alternation of optical lattice. c) Displacement of image in radian caused fluctuation of the imaging system. d) Photodiode signal for one lattice side of stabilized lattice in locked configuration..

Another stability issue is related to the fact, that measurement is not done in the place where atoms are located. To address this concern, we placed Thorlabs R1DS1P test target at the position where atoms were expected and illuminated it with one of the interfering laser beams. The data was recorded with the camera at a frame rate of 1500 Hz. The measured displacement of the test object image $\Delta$, in pixels, was converted into a phase shift for the smallest lattice period, $\Lambda_{min} = 12.82$ pixels, according to the relation $2\pi\Delta/\Lambda_{min}$. The displacement values are presented at the Figure 5c and demonstrate that the imaging system introduces noise with phase deviation $0.016$ rad which is much smaller than typical phase deviation of optical lattice $0.043$ rad and thus does not introduce any noticeable additional error to the lattice stability.

Besides feedback loop may add some high frequencies in the stabilized atomic cloud. In order to make sure the feedback loop does not have this issue one of the lattice side was imaged onto fast photodiode PDA05CF2 with bandwidth 150 MHz. The spectrum of diode data is presented at Figure 5d and does not seem to have any significant high frequency components.

## VI. CONCLUSIONS

Two possible geometries of the optical accordion were considered, with the configuration using a movable mirror found to be significantly more stable. In this configuration, the temperature dependence of the optical lattice phase drift was recorded. While a moderate correlation (0.5) between phase and temperature was observed, the overall system stability was influenced by additional factors. Analysis of the optical lattice phase noise spectrum revealed several resonance frequencies, which are likely excited through acoustic channels. Long-term stability was achieved using a feedback loop based on video camera input, enabling not only phase stabilization over extended periods but also full control of the optical lattice phase during rapid changes in lattice period. These developments may be applied to stabilize optical lattices in quantum simulation applications. The developed phase stabilization algorithm can be extended to more complicated 2D latices since it is purely relied on Fourier analysis of the lattice image and thus may easily handle much more complicated structures, than just periodic lattice.

## VII. ACKNOWLEDGMENTS

This work was supported by Rosatom in the framework of the Roadmap for Quantum computing (Contract No. 868-1.3-15/15-2021 dated October 5, 2021).

## AUTHOR DECLARATIONS

## CONFLICT OF INTEREST

The authors have no conflicts to disclose.

## AUTHOR CONTRIBUTIONS

**Pavel Aksentsev**: Conceptualization (equal), Data curation (equal), Formal analysis (equal), Investigation (lead), Methodology (equal), Software (equal), Visualization (equal), Writing – original draft (lead), Writing – review & editing (equal).

**Vladimir Khlebnikov**: Conceptualization (equal), Data curation (equal), Formal analysis (equal), Investigation (equal), Methodology (lead), Software (equal), Visualization (equal), Writing – original draft (equal), Writing – review & editing (equal).

**Ivan Cojocaru**: Software (lead).

**Arjuna Rudnev**: Investigation (supporting), Formal analysis (supporting), Methodology (equal).

**Ivan Pyrkh**: Formal analysis (supporting), Investigation (supporting).

**Davlet Kumpilov**: Formal analysis (supporting), Investigation (supporting).

**Polina Trofimova**: Formal analysis (equal), Validation (supporting).

**Ayrat Ibrahimov**: Visualization (equal).

**Oleg Blochin**: Formal analysis (supporting).

**Kirill Frolov**: Software (supporting).

**Sergey Kuzmin**: Software (supporting).

**Anna Zykova**: Resources (equal).

**Daniil Pershin**: Methodology (equal), Software (equal).

**Vladislav Tsyganok**: Conceptualization (lead), Investigation (supporting), Methodology (equal), Project administration (equal), Supervision (equal).

**Alexey Akimov**: Conceptualization (equal), Funding acquisition (lead), Project administration (lead), Resources (lead), Supervision (lead), Writing – review & editing (lead).

## DATA AVAILABILITY

The data that support the findings of this study are available from the corresponding author upon reasonable request.